\documentstyle[aps,prd,epsf]{revtex}
\draft
\newcommand{\f}[2]{{#1 \over #2}}
\newcommand{\Ref}[1]{(\ref{#1})}
\newcommand{\0}{{\hspace{-.5ex}\rm o}}
\newcommand{\1}{{\hspace{-.1ex}\rm o}}
\newcommand{\p}{{\hspace{-.5ex} p}}
\newcommand{\2}{{\hspace{-.1ex} p}}
\newcommand{\be}{\begin{equation}}
\newcommand{\ee}{\end{equation}}
\newcommand{\bn}{\begin{eqnarray}}
\newcommand{\en}{\end{eqnarray}}
\newcommand{\bd}{\begin{displaymath}}
\newcommand{\ed}{\end{displaymath}}
\newcommand{\bnn}{\begin{eqnarray*}}
\newcommand{\enn}{\end{eqnarray*}}
\newcommand{\bml}{\begin{mathletters}}
\newcommand{\eml}{\end{mathletters}}
\begin{document}
\title{Self-energy and Self-force in the Space-time of a Thick Cosmic String} 
\author{N.R. Khusnutdinov${}^{a)}$\thanks{e-mail:nail@dtp.ksu.ras.ru} and 
V.B. Bezerra${}^{b)}$\thanks{e-mail:valdir@fisica.ufpb.br}}
\address{a) Department of Theoretical Physics, Kazan State Pedagogical
University, \\Kazan, Mezhlauk 1, 420021, Russia\\
b) Departamento de F\'{\i}sica, Universidade Federal da
Para\'{\i}ba, \\
Caixa Postal 5008, CEP 58051-970 Jo\~ao Pessoa, Pb, Brazil}
\maketitle
\begin{abstract}
We calculate the self-energy and self-force for an electrically
charged particle at rest in the background of Gott-Hiscock cosmic
string space-time. We found the general expression for the self-energy which
is expressed in terms of the $S$ matrix of the scattering problem. The
self-energy continuously falls down outward from the string's center
with maximum at the origin of the string. The self-force is 
repulsive for an arbitrary position of the particle. It tends to zero in
the string's center and also far from the string and it has a maximum 
value at the string's surface. The plots of the numerical calculations of the
self-energy and self-force are shown.  
\end{abstract}
\section{Introduction}
Topological defects may have been formed during the
expansion of the Universe due to spontaneous symmetry breaking
\cite{Kibb76}. Among these defects, cosmic strings seem to be of particular
interest \cite{VileShel} due to its very intriguing properties compared
with those associated with a non-relativistic linear distribution of matter. 

A variety of models of strings has been proposed. The first model 
corresponding to an infinitely thin cosmic string has been considered by 
Vilenkin in Ref.\cite{Vile81}. This space-time is locally flat but globally 
it is the direct product of the two-dimensional Minkowski space-time and
a two-dimensional cone. The main geometrical peculiarity of this
space-time is the deficit angle. The Riemann tensor is a  delta-function with
support on the string's origin \cite{SokoStar77}. 

A more realistic model of a cosmic string with inner structure
has been considered by Gott \cite{Gott85} and Hiscock \cite{Hisc85}. The 
interior of this string is a constant curvature space-time and the exterior 
is a conical space-time as in the Vilenkin model. The deficit angle of the 
conical space is expressed in terms of the energy density of matter inside 
the string as we will see in Sec. \ref{Space-Time}. There is no singularity 
at the origin of the string and the Riemann tensor is constant inside the 
string and zero outside it. The metric functions are $C^1$-regular at the 
string's surface. 

The space-time of a cosmic string in the $U(1)$ gauge theory has been
considered numerically by Garfinkle in Ref.\cite{Garf85}. It was shown
that this space-time smoothly tends to a conical space-time far from the
string's origin. 

The self-energy and the self-force for (un)charged particle
in the infinitely thin cosmic string space-time has been investigated in
Refs.\cite{Line86,Smit90,Galt90,Khus94,Khus95}. The self-force on electric 
and magnetic linear sources in this space-time was also investigated 
in Ref.\cite{Bez1} and extended to the case of multiple cosmic 
strings\cite{Bez2}.

It is well-known\cite{Vile81} that a straight cosmic string produces no
gravitational force on surrounding matter. Nevertheless, there is 
interaction between the cosmic string and a particle due to its own field. 
This effect is non-local and depends on the deficit angle of the conical 
section of the cosmic string space-time. It was shown that even for 
a particle at rest, the electromagnetic force has only radial component, it 
is repulsive\cite{Line86,Smit90,Khus94,Khus95} and given by 
\be
F_r^{el} = L_0\f{q^2}{2r^2}\ .\label{el}
\ee
In the gravitational case it is attractive\cite{Galt90} and has the 
following expression 
\be
F_r^{gr} = -L_0\f{m^2}{2r^2}\ . \label{gr}
\ee
Here, $q$ and $m$ are charge and mass of particle, respectively; $r$ is
a distance from particle to string. 

The constant $L_0$ depends only on the deficit angle of the conical
section of the cosmic string space-time. It is zero for zero angle deficit
but for a supermassive cosmic string \cite{Smit90,Khus95} it becomes very
strong ($L_0\to\infty$).  

As it is seen from Eqs.\Ref{el} and \Ref{gr} the self-interaction force tends
to infinity at the string's origin, that is, as $r\to 0$. Obviously, real
cosmic strings which may appear at phase transitions in the early Universe
have non-zero thickness. For example, cosmic strings that appeared in Grand 
Unified Theory (GUT) have radius $r_\0\sim 10^{-29}cm$ \cite{VileShel}. In 
this case the space-time inside the string is not flat and one may expect 
some modification in the self-force that arises from this fact. This 
problem has been qualitatively discussed in Ref.\cite{PerkDavi91} in the 
context of the study of implications of cosmic string catalysis in the 
process of baryon decay.  

The purpose of this paper is to calculate the electromagnetic
self-energy and self-force for a charged particle at rest
in the background of Gott-Hiscock cosmic string space-time presented 
in Refs.\cite{Gott85,Hisc85}. In order to do the calculation we use the 
approach developed in Refs.\cite{Line86,Smit90}. In this method the
self-potential $\Phi$ and self-energy $U$ are proportional to the
coincidence limit at the particle position ${\vec x}_p$ of the renormalized
Green's function of the three-dimensional Laplace operator 
\bn
\Phi({\vec x}_p) &=& 4\pi qG^{ren}({\vec x}_p|{\vec x}_p)\ ,\\
U({\vec x}_p)&=\f 12q&\Phi({\vec x}_p)\ ,
\en
and the self-force is the minus gradient of self-energy 
\be
{\vec F}({\vec x}_p) = -\vec{\nabla}_{x_p} U({\vec x}_p)\ .
\ee
To renormalize the Green's function we use standard procedure 
\cite{BrowOtte86,Chri78} and subtract from that its singular part in
Hadamard's form.  

The organization of this paper is as follows. In Sec. \ref{Space-Time} we
introduce Gott-Hiscock space-time generated by a finite thickness cosmic string
space-time and describe all geometrical characteristics we need. In Sec.
\ref{Self} we obtain the general formulas for the self-potential for a 
particle at rest in this background taking into account the contributions
from inner and outer parts of the string.
In Sec. \ref{Discussion} we analyze qualitatively and numerically the
self-energy and self-force. We discuss our results in
Sec.\ref{Conclusions}. In Appendix \ref{A} we obtain an  uniform expansion 
for the self-energy and discuss the self-force. 

Throughout this paper we use units $c = G = 1$.

\section{The Space-Time}\label{Space-Time}
We consider the space-time of a straight infinite cosmic string with
constant energy density $\cal E$ corresponding to the matter inside it. The 
solution of Einstein equations for this case has been found by 
Gott\cite{Gott85} and Hiscock \cite{Hisc85}. The space-time may be divided 
by the surface of the string in two parts: interior and exterior domains. 
The latter is a flat conical space-time described by the line element 
\be
ds^2_{out}=-dt^2 + dr^2 + \f{r^2}{\nu^2}d\varphi^2 + dz^2\ ,\label{out}
\ee 
where $\varphi\in [0,2\pi ]$, $r\in [r_\0,\infty )$ and the parameter 
$r_\0$ is the radius of the string. 

The interior part of the string is a constant curvature space-time with 
line element given by
\be
ds^2_{in}=-dt^2 + d\rho^2 + \f{\rho_\1 ^2}{\epsilon^2}\sin^2\left(\f{\epsilon
\rho}{\rho_\1 }\right)d\varphi^2 + dz^2\ ,\label{in}
\ee 
where $\varphi\in [0,2\pi ]$ and $\rho\in [0,\rho_\1]$. The inner radius of 
the string is $\rho_\1 $ and it is related with  parameter 
$\epsilon$ by $\rho_\1 /\rho_* = \epsilon$, where $\rho_* = 
1/\sqrt{8\pi{\cal E}}$ is the "energetic" inner radius of the string. The 
 matching conditions at the surface of the string give the connection 
between inner $(\rho_\1 ,\epsilon)$ and outer $(r_\0,\nu)$ parameters of 
the string, which reads as 
\be
\f{r_\0}{\rho_\1 } = \f{\tan\epsilon}\epsilon\ ,\ \nu = \f 1{\cos\epsilon}\ .
\ee 
In this case the surface energy-momentum tensor is absent. 

The non-zero components of the Riemann and Ricci tensors and scalar 
curvature are the following 
\be
R^{\rho\varphi}_{\cdot\ \cdot}{}_{\rho\varphi} = \f{\epsilon^2}{\rho_\1
^2}\ ,\ R^\rho_\rho = R^\varphi_\varphi = \f{\epsilon^2}{\rho_\1 ^2}\ ,\ 
R=\f{2\epsilon^2}{\rho_\1 ^2}\ .
\ee
The space-time described by Eqs.\Ref{out} and \Ref{in} may be covered by 
one map using the continuation of the radial coordinate $r$ into the inner 
domain of  the string through the relation  
\be
\sin\left(\f{\epsilon\rho}{\rho_\1 }\right) =\f r{r_\0} \sin \epsilon\ .
\label{relation}
\ee
Therefore the line element can be written as\cite{AlleOtte90}
\be
ds^2=-dt^2 + P^2(r)dr^2 + \f{r^2}{\nu^2}d\varphi^2 + dz^2\ ,\label{both}
\ee
where the function $P(r)$ is given by
\be
P(r)=\left\{\begin{array}{ccl}\left(\nu^2 +\f{r^2}{r_\0^2}(1-\nu^2)\right)^{-
1/2}&,&r\leq r_\0\ , \\
1&,&r\geq r_\0\ .
\end{array}\right.
\ee
Sometimes we use the metric inside the string in the form given by 
Eq.\Ref{in} instead of Eq.\Ref{both}. Connection between them is obtained
from Eq.\Ref{relation}. 
\section{The Self-Energy}\label{Self}
The electromagnetic potential $A^\mu$ in the Lorenz gauge for a particle with 
trajectory $x^\mu = x^\mu(\tau)$ obeys the equations 
\be
g^{\alpha\beta}\nabla_\alpha \nabla_\beta A^\mu + R^\mu_\nu A^\nu = 
-4\pi J^\mu(x)= - 4\pi q\int u^\mu(\tau)\delta^4(x-x(\tau))
\f{d\tau}{\sqrt{-g}}\ .
\ee

It is possible, in principle, to obtain the self-force for an arbitrary 
trajectory of the particle using the same procedure of 
Refs.\cite{Khus94,Khus95}, but for simplicity we shall consider the
particle at the rest with trajectory
\be
x^0(\tau )=\tau\ ,\ x^1(\tau ) = r_\p\ ,\ x^2(\tau )=\varphi =0\ ,\ 
x^3(\tau )= z =0\ , \label{traj}
\ee 
where $r_\p$ is the radial position of the particle. Let us consider the 
equation for the zero component of the vector potential $A^\mu$. In the 
space-time of a cosmic string with metric \Ref{both}, it reads
\be
(-\partial^2_t + \triangle )A^0 = -4\pi J^0\ ,
\ee 
where $\triangle = g^{ik}\nabla_{i} \nabla_{k}$ is the three-dimensional 
Laplacian. For our space-time and trajectory given by Eq.\Ref{traj}, we get 
\be
\triangle A^0 = -\f{4\pi q}{\sqrt{g^{(3)}}}\delta (r-r_\p)\delta (\varphi) 
\delta (z)\ , 
\ee 
where $g^{(3)} = r^2P^2(r)/\nu^2$  is the determinant of the 
three-dimensional part of the metric. Therefore, $A^0$ is the scalar 
Green's function of the three-dimensional Laplacian multiplied by $4\pi q$
\be
A^0(x,\varphi ,z) = 4\pi q G(r,\varphi,z|r_\p,0,0)\ .
\ee 
The self-potential $\Phi$ and self-energy $U$, according to 
Refs.\cite{Line86,Smit90} are
\bn
\Phi(r_\p) &=& 4\pi qG^{ren}(r_\p,0,0|r_\p,0,0)\ ,\label{selfpotential}\\
U(r_\p)&=\f 12q&\Phi(r_\p)\ ,\label{selfenergy}
\en
where $G^{ren}$ is the renormalized Green's function. 

Since the self-energy depends only on the radial coordinate $r$, the
self-force will have only radial component given by
\be
F_{r_\p} = -\f d{d r_\p } U(r_\p)\ .\label{selfforce}
\ee

Now, let us find in a closed form the three-dimensional scalar 
Green's function $G(x, x')$  
\be
\triangle G(x ,x') = -\f 1{\sqrt{g^{(3)}}}\delta (r-r')\delta (\varphi -
 \varphi')\delta (z-z')\ .\label{Green}
\ee
In the space-time under consideration, Eq.\Ref{Green} turns into 
\be
\left(\f 1{r P(r)}\f\partial{\partial r} \f r{P(r)}\f\partial{\partial r} + 
\f{\nu^2}{r^2}\f{\partial^2}{\partial \varphi^2} + 
\f{\partial^2}{\partial z^2}\right)G(x ,x') = -\f\nu {r P(r)}\delta (r-r')
\delta (\varphi - \varphi') \delta (z-z')\ .   
\ee
Taking into account the cylindrical symmetry of the problem we may represent 
the Green's function in the following form
\be
G(x ,x')=\f 1{4\pi^2}\int_{-\infty}^\infty d k_{z} e^{ik_{z}(z-z')} 
\sum_{n=-\infty}^\infty e^{in(\varphi - \varphi')}\phi(r, r')\ , 
\ee
where the radial part of Green's function obeys the equation 
\be
\left(\f 1{r P(r)}\f\partial{\partial r} \f r{P(r)}\f\partial{\partial r} - 
\f{n^2\nu^2}{r^2} - k_z^2\right)\phi(r, r')=-\f{\nu}{r P(r)}\delta (r-r')\ . 
\label{radial}
\ee
In order to calculate the radial Green's function we use the standard approach 
and the following expression for it 
\be
\phi(r, r') = \theta (r-r')\psi_1(r)\psi_2(r')+
\theta (r'-r)\psi_1(r')\psi_2(r)\ ,
\ee 
where $\theta$ is the step function and $\psi_1$, $\psi_2$ are two linearly 
independent homogeneous solutions of Eq.\Ref{radial}. The function $\psi_1(r)$ 
falls  down as $r\to \infty$ and $\psi_2(r)$ is regular at the origin. 
Integrating Eq.\Ref{radial} over $r$ around $r'$ one has the Wronskian 
normalization condition \cite{AlleOtte90}
\be
W(\psi_2,\psi_1)=\psi_1'(r)\psi_2(r) - \psi_1(r)\psi_2'(r) = -\f\nu r P(r)\ .
\label{wronskian}
\ee
Both homogeneous solutions are  regular at the string's surface and 
they obey the set of equations
\bml\bn
\psi_{k}(r)_{|r_\0+\varepsilon}&=&\psi_{k}(r)_{|r_\0-\varepsilon}\ ,\\ 
\psi'_{k}(r)_{|r_\0+\varepsilon}&=&\psi'_{k}(r)_{|r_\0-\varepsilon}\ ,
\en \label{regular}\eml 
with $\varepsilon\to 0$ and $k=1,2$. 

The homogeneous solutions $\psi$ yield Bessel's equation 
\be
\left(\f 1r\f\partial{\partial r} r\f\partial{\partial r} - 
\f{n^2\nu^2}{r^2} - k_z^2\right)\psi_{out}(r) = 0\ ,
\ee
outside the string and Legendre's equation 
\be
\left(\f 1{\sin\left(\f{\epsilon\rho}{\rho_\1 }\right)}\f\partial{\partial 
\rho}
\sin\left(\f{\epsilon\rho}{\rho_\1 }\right)\f\partial{\partial \rho} - 
\f{n^2\epsilon^2}{\rho_0^2\sin^2\left(\f{\epsilon\rho}{\rho_\1 }\right)} - 
k_z^2\right)\psi_{in}(\rho) = 0\ ,
\ee
inside it. 

Therefore the homogeneous solutions with required boundary conditions are 
the following 
\be
\psi_1 =\left\{ \begin{array}{lll}
                 N\cdot p^{|n|}_\mu [x]+
                 M\cdot q^{|n|}_\mu [x]&;&r\leq r_\0\\
                 F\cdot K_{|n|\nu}[k r]&;&r\geq r_\0
                \end{array}
        \right. \ ,\ 
\psi_2 =\left\{ \begin{array}{lll}
                 E\cdot p^{|n|}_\mu [x]&;&r\leq r_\0\\
                 A\cdot I_{|n|\nu}[k r]+B\cdot K_{|n|\nu}[k r]&;&r\geq r_\0
                \end{array}
        \right. \ ,\label{HomSol}
\ee
where $k=|k_{z}|$ and $x=\cos\left(\f{\epsilon\rho}{\rho_\1 }\right)=
(\nu P(r))^{-1}$; $I_{n\nu}$ and $\ K_{n\nu}$ are modified Bessel
functions. The functions $p^{|n|}_\mu$ and $q^{|n|}_\mu$ are expressed in
terms of the Legendre functions of first and second kind by  
\bn  
p^{|n|}_\mu[x]&=&P^{-|n|}_\mu[x]\ ,\\
q^{|n|}_\mu[x]&=&\f{(-1)^n}2\left(Q^{n}_\mu[x]+Q^{n}_{-\mu-1}[x] \right)
\nonumber\\ 
&=&-\f\pi{2\sin\pi\mu}P^{|n|}_\mu[-x]\ ,
\en
where $\mu = -\f 12 +\f 12\sqrt{1-4k^2r_\0^2/(\nu^2-1)}$.

On choosing these functions, they will be real for arbitrary value of 
$x$ and the Wronskian of them will take the simple form 
\be
W(p^{|n|}_\mu, q^{|n|}_\mu)=\f 1{1-x^2}\ . 
\ee

The six constants in Eq.\Ref{HomSol} can be found from  
Eqs.\Ref{wronskian} and \Ref{regular}. The Wronskian normalization condition 
Eq.\Ref{wronskian} gives two relations 
\be
AF=\nu\ ,\ ME=1\ ,
\ee
and the conditions of regularity on the surface of the string given by
Eq.\Ref{regular} define the ratios
\bml\bn
S_{out}[kr_\0,|n|,\nu]&\stackrel{def}{=}&\f BA =
-\f{\nu\left(1-\f 1{\nu^2}\right)p^{|n|}_\mu{}'[\f
1\nu]I_{|n|\nu}[kr_\0]+kr_\0 p^{|n|}_\mu [\f 1\nu] 
I'_{|n|\nu}[kr_\0]}{\nu\left(1-\f 1{\nu^2}\right)
p^{|n|}_\mu{}'[\f 1\nu]K_{|n|\nu}[kr_\0]+kr_\0 p^{|n|}_\mu [\f 1\nu]
K'_{|n|\nu}[kr_\0]}\ ,\\ 
S_{in}[kr_\0,|n|,\nu]&\stackrel{def}{=}&\f NM =
-\f{\nu\left(1-\f 1{\nu^2}\right) 
q^{|n|}_\mu{}'[\f 1\nu]K_{|n|\nu}[kr_\0]+kr_\0 q^{|n|}_\mu [\f 1\nu]
K'_{|n|\nu}[kr_\0]}{\nu\left(1-\f 1{\nu^2}\right)
p^{|n|}_\mu{}'[\f 1\nu]K_{|n|\nu}[kr_\0]+kr_\0 p^{|n|}_\mu [\f 1\nu]
K'_{|n|\nu}[kr_\0]}\ .
\en\eml
These quantities characterize the scattering on the string. The $S$-matrix 
$S_{out}$ may be represented in the following form $S_{out}=f_{n}^*(ik)/
f_{n}(ik)$, where $f_{n}(ik)$ is the Jost function of the scattering problem
on the imaginary axis. It has been found in Ref. \cite{KhusBord99} and obeys
the relations 
\bn
f_{n}(i k)&=& -\f 1{\sqrt{\nu}}\left(\f{kr_\0}{\sqrt{\nu^2-1}}\right)^n
\left\{\nu\left(1-\f 1{\nu^2}\right)p^{|n|}_\mu{}'[\f 1\nu]K_{|n|\nu}[kr_\0]+
kr_\0 p^{|n|}_\mu [\f 1\nu]K'_{|n|\nu}[kr_\0]\right\}\ ,\label{Jost}\\
f_{n}^*(i k)&=& +\f 1{\sqrt{\nu}}\left(\f{kr_\0}{\sqrt{\nu^2-1}}\right)^n
\left\{\nu\left(1-\f 1{\nu^2}\right)p^{|n|}_\mu{}'[\f 1\nu]I_{|n|\nu}[kr_\0]+
kr_\0 p^{|n|}_\mu [\f 1\nu]I'_{|n|\nu}[kr_\0]\right\}\ .\label{Jost*}
\en 

Therefore the radial Green's function of our problem is the following ($r>r'$)
\be
\phi(r, r')=\left\{\begin{array}{lll}\nu K_{|n|\nu}[k r']\left(I_{|n|\nu}[k r]+
S_{out}[kr_\0,|n|,\nu]K_{|n|\nu}[k r]\right)&;&r, r'\geq r_\0\\
p^{|n|}_\mu [x']\left(q^{|n|}_\mu [x]+S_{in}[kr_\0,|n|,\nu ]
p^{|n|}_\mu [x]\right)&;&r, r'\leq r_\0\end{array}\right.\ ,
\ee
where $x=\cos\left(\f{\epsilon\rho}{\rho_\1 }\right)=
(\nu P(r))^{-1}$ and $x'=\cos\left(\f{\epsilon\rho'}{\rho_\1 }\right)=
(\nu P(r'))^{-1}$. In the limit $\nu\to 1$ this function becomes 
\be
\phi (r, r') =  K_{|n|}[k r'] I_{|n|}[k r]\ , 
\ee 
as it should be in flat space-time. 

Let us proceed now to calculate the self-energy given by Eq.\Ref{selfenergy}. 
The zero component of the vector potential $A^0$ of a particle with trajectory 
given by Eq.\Ref{traj} and situated outside the string is 
\be
A^0(r,\varphi, z)=\f{q\nu}{\pi}\int_{-\infty}^\infty d k_{z} e^{ik_{z} z} 
\sum_{n=-\infty}^\infty e^{in\varphi}K_{|n|\nu}[k r_\p]
\left(I_{|n|\nu}[k r]+S_{out}[kr_\0,|n|,\nu]K_{|n|\nu}[k r]\right) \ . 
\ee
Taking the coincidence limit in this expression for a fixed value of the 
angular variable, for example, $\varphi = 0$, and changing the integration 
variable 
$k_{z}\to k=|k_{z}|$, one has 
\be
A^0(r, z)=\f{2q\nu}\pi\int_0^\infty d k \cos (k z) 
\sum_{n=-\infty}^\infty K_{|n|\nu}[k r_\p]
\left(I_{|n|\nu}[k r]+S_{out}[kr_\0,|n|,\nu]K_{|n|\nu}[k r]\right) \ . 
\ee
This expression consists of two parts. The first one is due to the first 
term in brackets which is exactly the potential for an infinitely thin 
cosmic string. This term is divergent in the coincidence limit 
$z\to 0\ ,\ r\to r_\p$. The second term is finite in this limit and tends 
to zero as $\nu\to 1$ as well as $r_\0 \to 0$. 
Therefore to renormalize this potential we have to renormalize just the 
first term. Because the exterior is a flat space-time, in order to do the
renormalization we may subtract from it the potential in Minkowski space-time 
which corresponds to $\nu =1$. 
Therefore the self-potential has the following form
\bn
\Phi(r_\p)&=&\lim_{z\to 0}\f {2q}{\pi}\int_0^\infty d k\cos (k z) 
\sum_{n=-\infty}^\infty\left(\nu K_{|n|\nu}[k r_\p]I_{|n|\nu}[k r_\p] - 
 K_{|n|}[k r_\p]I_{|n|}[k r_\p]\right)\nonumber\\
&+&\f{2q\nu}{\pi}\int_0^\infty d k 
\sum_{n=-\infty}^\infty S_{out}[kr_\0,|n|,\nu]K^2_{|n|\nu}[k r_\p] \ . 
\en
The first contribution may be found in closed form \cite{Line86,Smit90} using 
formulas 6.672(3) and 8.715(2) from Ref.\cite{GR} and we arrive at the 
following expression for the self-potential for a particle in the exterior of 
the string ($R=r_\p/r_\0$)
\be
\Phi (r_\p)=\f q{r_\p} L(\nu, R)\ ; \ R\geq 1 ,
\label{Phiout}
\ee
where
\be
L(\nu, R)= \f 1\pi\int_0^\infty \f{\nu\coth (\nu x) - 
\coth x}{\sinh x} d x + \f{2\nu}{\pi}\int_0^\infty d x 
\sum_{n=-\infty}^\infty S_{out}[\f x R, |n|,\nu]K^2_{|n|\nu}[x] \ . 
\ee
This formula represents an interesting relation between the self-potential 
with the scattering problem and the Jost function on this background. The 
first term is a well-known result \cite{Line86,Smit90} for an infinitely 
thin string. The second term is the contribution due to non-zero thickness 
of the string. It tends to zero as the radius of the string goes to 
zero ($r_\0\to 0$). 

Let us now consider a particle situated in the interior of the string. In 
this case the zero component of the vector potential reads  
\be
A^0(r,\varphi, z)=\f {2q}\pi\int_0^\infty d k \cos k z 
\sum_{n=-\infty}^\infty e^{in\varphi}p^{|n|}_\mu [x']
\left(q^{|n|}_\mu [x]+S_{in}[kr_\0,|n|,\nu ]p^{|n|}_\mu [x]\right) \ . 
\label{selfinterior}
\ee
To renormalize this expression we have to subtract from it all divergences in 
the Hadamard form \cite{BrowOtte86}. The structure of divergences of Green's 
function for odd-dimensional spaces is more simple then for 
even-dimensional case because in the former case there is no logarithmic 
singularity \cite{Chri78}. 
The singular part of the Green's function in three dimensions is\cite{Chri78} 
\be
G_{sing}(x, x')=\f{\triangle^{1/2}}{4\pi}\f 1{\sqrt{2\sigma}}\ .
\ee
Taking the coincidence limit for angular variable $\varphi$ we get
\be
G_{sing}(\rho, z |\rho', z')=\f{\triangle^{1/2}}{4\pi}\f 1{
\sqrt{(\rho - \rho')^2 + (z-z')^2}}\ , 
\label{singular}
\ee
where 
\be
\triangle^{1/2} = 1 + \f 1{12}\f{\epsilon^2}{\rho_\1 ^2}(\rho -\rho')^2 + 
\f 1{160}\f{\epsilon^4}{\rho_\1 ^4}(\rho -\rho')^4 + \cdots \ ,
\ee
and $2\sigma = (\rho -\rho')^2 + (z-z')^2$. Let us represent the singular
part of Green's function \Ref{singular} in the following integral form  
\be
G_{sing}(\rho, z|\rho', z') = \f{\triangle^{1/2}}{2\pi^2}
\int_0^\infty d k \cos k(z-z') \sum_{n=-\infty}^\infty 
K_{|n|}[k\rho] I_{|n|}[k\rho']\ . \label{Gsing}
\ee 
To renormalize the self-potential we subtract from Eq.\Ref{selfinterior} the 
above expression multiplied by $4\pi q$ and take the coincidence limit 
$\rho = \rho' = \rho_\2\ ,\ z=z'$. So, we arrive at the result
\be
\Phi(r_\p)=\f{2q}\pi\int_0^\infty d k 
\sum_{n=-\infty}^\infty\left\{ p^{|n|}_\mu [x_{p}]
\left(q^{|n|}_\mu [x_{p}]+S_{in}[kr_\0,|n|,\nu ]p^{|n|}_\mu [x_{p}]\right) - 
K_{|n|}[k\rho_\2] I_{|n|}[k\rho_\2] \right\} \ , 
\label{Phiin}
\ee
where 
\bn 
\mu &=& -\f 12 + \f 12\sqrt{1 - \f{4k^2r_\0^2}{\nu^2-1}}= 
 -\f 12 + \f 12\sqrt{1 - \f{4k^2\rho_\1 ^2}{\epsilon^2}}\ ,\\
x_{p}&=&(\nu P(r_\p))^{-1}=\cos \left(\f{\epsilon\rho_\2}{\rho_\1
}\right)\ , \ 
kr_\0= k\rho_\1 \f{\tan\epsilon}{\epsilon}\ .  
\en
Note that we can easily show that 
\bn
&&\lim_{z\to z'}\lim_{\rho\to\rho'}\lim_{r\to r'} \int_0^\infty d k \cos
k(z-z') \sum_{n=-\infty}^\infty\left\{ K_{|n|}[k\rho]
I_{|n|}[k\rho']-K_{|n|}[kr] I_{|n|}[kr']\right\}\label{Prop}\\
&&=\int_0^\infty d k \sum_{n=-\infty}^\infty\left\{ K_{|n|}[k\rho]
I_{|n|}[k\rho]-K_{|n|}[kr] I_{|n|}[kr]\right\}=0\ . \nonumber 
\en
This is due to the fact that the singular part of Green's function given by 
Eqs.\Ref{singular} and \Ref{Gsing} in coincidence limit $\rho'=\rho$ 
does not depend on $\rho$. It is simply given by $1/4\pi |z-z'|$.
For this reason we may change $\rho_\2$ and $r_\p$ in the last term in 
Eq.\Ref{Phiin} and the self-potential $\Phi$ given by Eqs.\Ref{Phiout} and 
\Ref{Phiin} is continuous at the string's surface. 
\section{Discussion}\label{Discussion}
From previous results we have the following expressions for the self-energy of 
charged particles at the point $R=r_\p/r_\0$ in the thick cosmic string
space-time 
\be
U(r_\p)=\f{q^2}{2r_\0}\left\{\begin{array}{ll}
\f{1}{R}L(\nu, R)&R\geq 1\\
\ \ H(\nu, R)&R\leq 1\end{array}
        \right.\ , 
\label{U}\ee
where 
\bn
L(\nu, R) &=& \f 1\pi\int_0^\infty \f{\nu\coth (\nu x) - 
\coth x}{\sinh x} d x \nonumber \\
&+& \f{2\nu}{\pi}\int_0^\infty d x \sum_{n=-\infty}^\infty S_{out}[\f x R,
|n|,\nu]K^2_{|n|\nu}[x]\ ,\label{L}\\ 
H(\nu, R) &=& \f{2}{\pi}\int_0^\infty d x 
\sum_{n=-\infty}^\infty\left\{ p^{|n|}_{\mu_{x}}[x_{R}]
\left(q^{|n|}_{\mu_{x}} [x_{R}]+S_{in}[x, |n|,\nu ]p^{|n|}_{\mu_{x}}
[x_{R}]\right)\right.\nonumber \\ 
&-&\left. K_{|n|}[xR] I_{|n|}[xR]\right\}\ ,\label{H}\\
S_{out}[x, |n|,\nu]&=&-\f{\nu\left(1-\f 1{\nu^2}\right) p^{|n|}_{\mu_{x}}{}'
[\f 1\nu]I_{|n|\nu}[x] + x p^{|n|}_{\mu_{x}} [\f 1\nu]
I'_{|n|\nu}[x]}{\nu\left(1-\f 1{\nu^2}\right) p^{|n|}_{\mu_{x}}{}'[\f
1\nu]K_{|n|\nu}[x] + x p^{|n|}_{\mu_{x}} [\f 1\nu] K'_{|n|\nu}[x]}\ ,
\label{Sout}\\  
S_{in}[x, |n|,\nu]&=&-\f{\nu\left(1-\f 1{\nu^2}\right) 
q^{|n|}_{\mu_{x}}{}'[\f 1\nu]K_{|n|\nu}[x]+ x q^{|n|}_{\mu_{x}}{}'[\f 1\nu]
K'_{|n|\nu}[x]}{\nu\left(1-\f 1{\nu^2}\right)
p^{|n|}_{\mu_{x}}{}'[\f 1\nu]K_{|n|\nu}[x]+ x p^{|n|}_{\mu_{x}} [\f 1\nu]
K'_{|n|\nu}[x]}\ ,\label{Sin}
\en
and we have introduced the following notations 
\bn
\mu_{x} &=& -\f 12 + \f 12\sqrt{1 - \f{4x^2}{\nu^2-1}}\ ,\\
x_{R} &=& (\nu P(r_\p))^{-1} = \sqrt{1 - R^2(1-\f 1{\nu^2})}\ .    
\en 

Let us analyze qualitatively the above expressions for self-energy. First
of all let us consider the particle situated outside the string. The function
$L(\nu, R)$ defined by Eqs.\Ref{U}, \Ref{L} can be separated into two parts
according to Eq.\Ref{L} as 
\be
L(\nu, R)=L_0(\nu) + L_1(\nu, R)\ , \label{LGen}
\ee
where  
\be
L_0(\nu)=\f 1\pi\int_0^\infty \f{\nu\coth (\nu x) - 
\coth x}{\sinh x} d x  
\ee
is the contribution to the self-energy due to the infinitely thin cosmic
string \cite{Line86,Smit90}. The second term, 
\be
L_1(\nu, R)=\f{2\nu}{\pi}\int_0^\infty d x 
\sum_{n=-\infty}^\infty S_{out}[\f x R, |n|,\nu]K^2_{|n|\nu}[x]\
,\label{Lr0} 
\ee
is the contribution from the structure of the string. 

The function $ S_{out}[\f x R, |n|,\nu] $ is positive for arbitrary angular
momentum $n$. For this reason the additional contribution to
self-energy due to non-zero thickness of the string is positive, too.
Changing the variable of integration $x\to z$ such that $x = n\nu z$ 
(except $n=0$) we can represent the function $L_1(\nu, R)$ as

\be
L_1(\nu, R)=\int_0^\infty d z F_0(z, R) + 
2\sum_{n=1}^\infty \int_0^\infty d z F_{n}(z, R)\ ,
\ee
where 
\bn
F_0(z, R)&=&\f{2\nu}{\pi}S_{out}[\f zR,0,\nu]K^2_0[z]\ ,\\
F_{n}(z, R)&=&\f{2n\nu^2}{\pi}S_{out}[\f z R n\nu, n,\nu]K^2_{n\nu}[z n\nu]\ .
\en
In order to estimate the behavior of $F_{n}$ as a function of $z$ we use the
uniform asymptotic expansion for great index $n$ of Bessel's functions in
Ref\cite{AbraSteg} and Legendre's function in Refs\cite{Torn,BordKhus}.
With the help of those expansions one obtains the following main term of
the uniform expansion for $F_{n}$ 
\be
F_{n}(z, R)\sim \f{\nu^2 -1}{8\pi \nu n^2
R^2}\f{z^2}{\sqrt{1+z^2}(1+\f{z^2}{R^2})^2}\exp \left\{-2n\nu \left(\eta
[z]-\eta [\f z R]\right)\right\}\ ,\label{Fn}  
\ee
where 
\be
\eta [z] = \ln\f z{\sqrt{1+z^2}+1} +\sqrt{1+z^2}\ .
\ee
The function in Eq.\Ref{Fn} tends to zero as $z^2$ for $z\to 0$ and it
tends to zero as $\exp \left\{-2n\nu z\left(1-\f
1{R^2}\right)\right\}/z^3$ for $z\to \infty$. For this reason the function
$L_1(\nu, R)$ exponentially falls down for great distance from the string
$R=r_\p/r_\0 \gg 1$ and it tends 
to a positive constant at the surface of the string at $R=1$. Therefore the
self-energy tends to that values for an infinitely thin cosmic string far 
from it. The main difference appears near the surface of the string where 
one has an additional positive contribution. 

The self-energy at string's origin may be analyzed by formulas
\Ref{U} and \Ref{H}. Taking the limit $R\to 0$ and using the behavior of
Legendre's function in the neighborhood of unit\cite{BateErde} one has the 
following expression for the self-energy in the origin
\be
U_{max}=\f{q^2}{2r_\0}\f 2\pi\int_0^\infty dx\left\{S_{in}[x,0,\nu]+
\f 12\ln\f{x^2}{\nu^2-1}-\Psi[\mu_x+1]-\f\pi 2\cot\pi\mu_x\right\}\ ,
\label{Umax}
\ee 
which is, in fact, the height of the potential barrier. Here, $\Psi$ is
the logarithmic derivative of the gamma function and 
\be
S_{in}[x,0,\nu]=-\f{\sqrt{\nu^2-1}}x\f{x K_1[x]q^0_{\mu_x}[\f 1\nu] -
\sqrt{\nu^2-1} K_0[x]q^1_{\mu_x}[\f 1\nu]}{\sqrt{\nu^2-1}
K_1[x]p^0_{\mu_x}[\f 1\nu] + x K_0[x]p^1_{\mu_x}[\f 1\nu]}\ . 
\ee 

As it can be seen from the uniform expansion Eq.\Ref{Fn}, the dependence of 
the self-energy on the metric coefficient $\nu=1/\cos\epsilon$ is given
mainly by the expression $(\nu - \f 1\nu)$. For this reason we represent the
self-energy $U$ and the height of the barrier $U_{max}$ as 
\bn
U&=&\f{q^2}{2r_\0}\f{\nu^2-1}\nu{\cal U}(\nu,R)\ ,\label{Ucal}\\
U_{max}&=&\f{q^2}{2r_\0}\f{\nu^2 - 1}\nu{\cal U}_{max}(\nu)\ ,
\label{Umax1} 
\en
where 
\be
{\cal U}_{max}(\nu)=\f\nu{\nu^2-1}\f 2\pi\int_0^\infty
dx\left\{S_{in}[x,0,\nu]+ \f 12\ln\f{x^2}{\nu^2-1}-\Psi[\mu_x+1]-\f\pi
2\cot\pi\mu_x\right\}\ . 
\ee

\begin{figure}
\centerline{\epsfxsize=8truecm\epsfbox{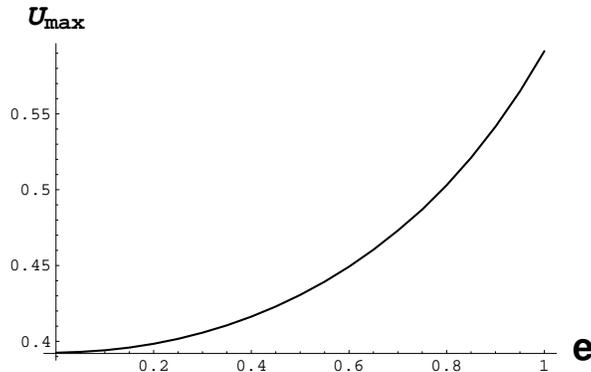}}
\caption{The self-energy at the string's origin ( height of the
potential barrier ) is represented as $U_{max} = \f{q^2}{2r_\0}
\f{\nu^2 -1}\nu {\cal U}_{max}(\nu)$. Here is the plot of the
dimensionless self-energy ${\cal U}_{max}$ as a function of
$\epsilon$. The cone parameter $\nu =1/\cos\epsilon$. For
$\epsilon\leq 0.1$ the ${\cal U}_{max}\approx 0.39$. \label{Fig1}}   
\end{figure}

The dependence of ${\cal U}$ on $\nu$ is weak for $\nu$ close to unit  
and it does not depend, in fact, on $\nu$ for $\epsilon\leq 0.1$. In
Fig.\ref{Fig1}, ${\cal U}_{max}(\nu)$ is displayed as a function of $\nu$.
 
Therefore for small deficit angle we obtain the following formula for
the height of the barrier 
\be
U_{max}\approx 0.39\f{q^2}{2r_\0}(\nu - \f 1\nu)\ .\label{Umaxapp}
\ee
The numerical calculation of ${\cal U}(\nu ,R)$ as a 
function of $R=r_\p/r_\0$ for $\epsilon = 0.1$ is depicted in
Fig.\ref{Fig2} (See Appendix A for details). 

\begin{figure}
\centerline{\epsfxsize=8truecm\epsfbox{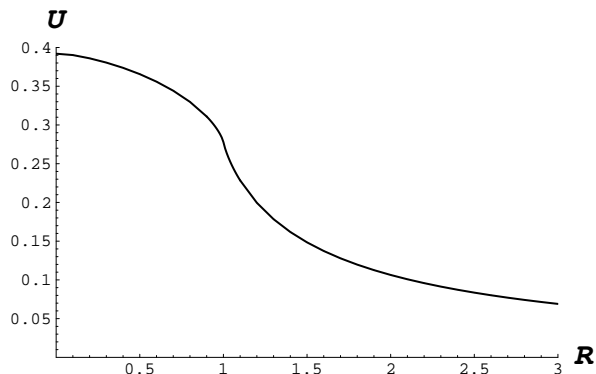}}
\caption{The self-energy $U(\nu, R) = \f{q^2}{2r_\0} \f{\nu^2 -1}\nu
{\cal U}(\nu, R)$. This is the plot of a dimensionless self-energy ${\cal
U}$ as a function of the particle position $R=r_\p/r_\0$ for
$\epsilon = 0.1$.  \label{Fig2}} 
\end{figure} 

Let us now compare the self-energy in the thick cosmic string space-time
with that in the infinitely thin cosmic string space-time in the limit
of zero thickness of the string ($r_\0\to 0$). We have to compare the
self-energy in two different space-times for a particle situated at the same
proper distance from the string. In the infinitely thin cosmic string
background the distance from the string $d$ coincides with the coordinate of
the particle, that is, $d=r_\p=r_\0R$. In the Gott-Hiscock cosmic string 
space-time the distance from the string to a particle is 
\be\label{PD}
d=\left\{\begin{array}{ll}
r_\0\left(R - 1+\f\epsilon{\tan\epsilon}\right)\ ,&R\geq 1\\ 
r_\0\f{\arcsin (R\sin\epsilon)}{\tan\epsilon}\ ,&R\leq 1
         \end{array}
   \right.\ , 
\ee
Taking into account the above expression and formulas Eqs.\Ref{U} and 
\Ref{L} we get for a particle situated outside the string $(D=d/r_\0)$,
the following relation 
\be
\f{U_{thick}}{U_{inf.thin}} =\f{D}{D+1-\f\epsilon{\tan\epsilon}}\left\{1+
\f{L_1(\nu,D+1-\f\epsilon{\tan\epsilon})}{L_0(\nu )}\right\}\ . 
\ee
In the limit of zero thickness $r_\0\to 0\ (D\to\infty)$ we obtain unit in
the rhs and the self-energy in the Gott-Hiscock space-time tends to the same 
result obtained in the infinitely thin cosmic string space-time. In this 
case the barrier's height $U_{max}$ given by Eq.\Ref{Umax1} tends to 
infinity $\sim 1/r_\0$.  

\begin{figure}             
\centerline{\epsfxsize=8truecm\epsfbox{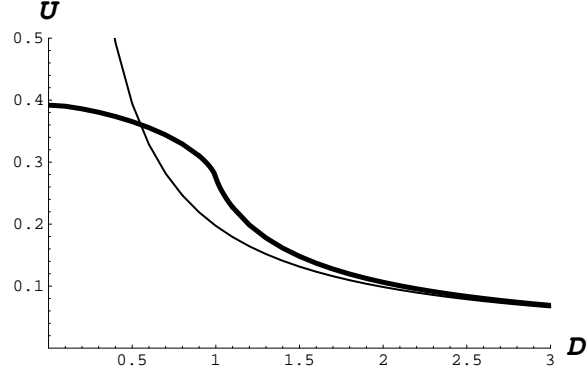}}
\caption{The dimensionless self-energy ${\cal U}$ (thick line) for a
particle in a thick cosmic string space-time and for a particle in an
infinitely thin cosmic string space-time ${\cal U}_{thin}$ (thin line) as
a function of the proper distance $D = d/r_\0$ numerically calculated for 
$\epsilon = 0.1 $. \label{Fig3}} 
\end{figure}

In Fig.\ref{Fig3} we display the numerical calculation of the self-energy
${\cal U}$ given by Eq.\Ref{Ucal} for a particle in the Gott-Hiscock
space-time and in an infinitely thin cosmic string space-time ${\cal
U}_{inf.thin}$ defined below 
\bn
U_{inf.thin}&=&\f{q^2}{2r_\0}\f{\nu^2-1}\nu{\cal U}_{inf.thin}\ , \\
{\cal U}_{inf.thin}&=&\f\nu{\nu^2-1}\f 1RL_0(\nu)\nonumber
\en
as a function of distance from the string's origin $D=d/r_\0$. 

For an infinitely thin cosmic string space-time $R=D$ and for the Gott-Hiscock
space-time we have
\be
R=\left\{\begin{array}{ll}
D + 1 -\f\epsilon{\tan\epsilon}\ ,&R\geq 1\\ 
\f{\sin(D\tan\epsilon)}{\sin\epsilon}\ ,&R\leq 1\end{array}
        \right.\ . 
\ee  

The self-force, according to Eq.\Ref{selfforce} is minus the derivative of
the self-energy given by Eq.\Ref{U} with respect to the particle position. 
It is zero in the string's origin and it tends to zero such as in the 
infinitely thin cosmic string space-time far away from the string. In the 
neighborhood of the string's surface, for $|R-1|\ll 1$, it tends to infinity 
logarithmically according to (see Appendix A) 
\be
F_r\approx -\f{q^2}{2r_\0^2}\f{\nu^2-1}8\ln |R-1|\
,\label{Fdiv} 
\ee
This divergence is rather formal. The function $\ln x$ is an integrable 
function at $x=0$. For this reason the work against this self-force is 
finite and equal to the height of the barrier $U_{max}$ given by 
Eq.\Ref{Umax1}.  

\begin{figure}             
\centerline{{\epsfxsize=8truecm\epsfbox{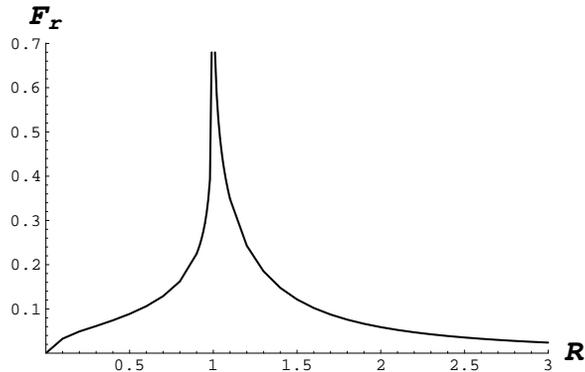}}}
\caption{The self-force $F_r(\nu, R) = \f{q^2}{2r_\0} \f{\nu^2 -1}\nu
{\cal F}_r(\nu, R)$. Here is the plot of the dimensionless
self-force for a particle in a thick cosmic string space-time
${\cal F}_r$ as a function of $R=r_\p/r_\0$ for $\epsilon = 
0.1$. \label{Fig4}}
\end{figure}

Because the self-potential $U$ has the structure given in Eq.\Ref{Ucal}
we represent the self-force in the same way as 
\be
F_r=\f{q^2}{2r_\0^2}\f{\nu^2-1}\nu{\cal F}_r(\nu, R)\ .\label{Fcal}
\ee
The numerical simulation of the function ${\cal F}_r$ is displayed in
Fig.\ref{Fig4}.  

\section{Conclusions}\label{Conclusions}
The aim of the paper is to calculate the self-energy and self-force for
a charged particle at rest in the space-time of an infinitely long,
straight cosmic string with non-trivial internal structure. The relevance
of this calculation is to clear up the role of the non-zero thickness of 
the string. It is well-known that for a particle at rest in the infinitely 
thin cosmic string space-time, the self-energy and self-force fall down 
far from the string and tends to infinity at the string's 
core\cite{Line86,Smit90}. Obviously the origin of these singularities are 
associated with the delta-like model of string's interior. 

In the proceeding sections we considered the self-energy and the self-force
of charged particles at rest in the cosmic string space-time with simplest
non-trivial interior, suggested by Gott \cite{Gott85} and Hiscock
\cite{Hisc85}. This model of string is an exact solution of the Einstein
equations and it corresponds to the cylindrical distribution of matter of
constant energy density inside it. The exterior of the string is the flat 
conical space-time and the interior is the constant curvature space-time. 
The model is usually named as "ballpoint pen" model \cite{AlleOtte90}. It is 
suitable for our goal because this model contains a dimensional parameter - 
the radius of string $r_\0$ - with respect of which we may analyse our 
problem. Let us summarise main results in what follows. 

To calculate the self-energy and self-force we used the approach of
Refs.\cite{Line86,Smit90} in which the quantities under consideration are
expressed in terms of the renormalized Green's function of the
three-dimensional Laplace operator. 

We analyzed the self-energy and self-force for different positions $r_\p$ of 
the particle . The self-energy falls down outside the string and 
tends to the same result obtained in the case of an infinitely thin 
cosmic string space-time far from string's surface. Namely, the self-energy 
has the following structure 
\be
U(r_\p)=\f{q^2}{2r_\p}\left[L_0(\nu) + L_1(\nu, \f{r_\p}{r_\0})\right]\ ,
\ee
where the function $L_1$ exponentially tends to zero far away from the string
and the function $L_0$ is the same one that corresponds to the case of an 
infinitely thin cosmic string space-time. The additional contribution to the 
self-energy is expressed as momentum expansion in terms of the $S$ matrix 
of the scattering problem in the imaginary axis.  

Inside the string the self-energy grows up and tends to a constant in the
string's origin, which is, in fact, the height of the potential barrier.
For a cosmic string which is considered in grand unified theory with $\nu -1
\approx 10^{-6}$ and $r_\0 \approx 10^{-29} cm$, the height of the energy
barrier is $2.8\cdot 10^5\ GeV$. 

In the limit of zero radius of the string ($r_\0\to 0$), the self-potential
tends to the same value corresponding to a particle in the infinitely 
thin cosmic string space-time and the maximum of the self-energy tends to 
infinity as $1/r_\0$.  

The self-force, which is the minus gradient of the self-energy
has only radial component and it is repulsive for any position of the
particle. It is zero in the string's origin and tends to the self-force in the
infinitely thin cosmic string space-time. In the surface of the string it
has a maximum value. In the framework of Gott-Hiscock thick cosmic string 
space-time the self-force tends to infinity logarithmically. This is an
integrable divergence and the total work against the self-force is finite
and equal to maximum of the self-energy at the center of the string. 

Therefore, the non-zero radius of the string drastically changes the
bahaviour of self-energy and self-force close to string's core. Outside
the string's surface the additional contributions due to string's radius
exponentially fall down. The self-energy and self-force are equal, in fact, 
to that in an infinitely thin cosmic string space-time starting from the
distance of two radius of the string. We expect that this bahaviour of
self-energy and self-force will be, in general, the same for a cosmic string
in the Abelian Higgs model, considered in \cite{Garf85}.

\acknowledgments 
We would like to thank Dr. M. Bordag for a critical reading of this 
paper. NK is grateful to Departamento de F\'{\i}sica, Universidade
Federal da Para\'{\i}ba (Brazil) where this work was done, for hospitality.
This work was supported in part by CAPES and in part by the Russian Found
for Basic Research, grant No 99-02-17941. VBB also would like to thanks 
Conselho Nacional de Desenvolvimento Cientifico e Tecnol\'ogico (CNPq) for 
partial financial support.

\appendix
\section{}\label{A}
In this section we discuss the numerical analysis of the self-potential given
by Eq.\Ref{U}. Using Eq.\Ref{Prop} we can represent the
self-potential in the following form 
\be
U(r_\p)=\f{q^2}{2r_\0}\left\{\begin{array}{ll}
\f{1}{R}L(\nu, R)&R\geq 1\\
\ \ H(\nu, R)&R\leq 1\end{array}
        \right.\ , 
\ee
where 
\bn
L(\nu, R) &=& \f 2\pi\int_0^\infty dx \sum_{n=-\infty}^{+\infty}\left\{\nu
K_{|n|\nu}[x]I_{|n|\nu}[x]-K_{|n|}[\f x\nu]I_{|n|}[\f x\nu]\right.
\nonumber \\ 
&+&\left. \nu S_{out}[\f x R, |n|,\nu]K^2_{|n|\nu}[x]\right\}\ ,\\ 
H(\nu, R) &=& \f{2}{\pi}\int_0^\infty d x 
\sum_{n=-\infty}^\infty\left\{ p^{|n|}_{\mu_{x}}[x_{R}]
\left(q^{|n|}_{\mu_{x}} [x_{R}]+S_{in}[x, |n|,\nu ]p^{|n|}_{\mu_{x}}
[x_{R}]\right)\right. \nonumber\\ 
&-&\left. K_{|n|}[\f{x R}\nu] I_{|n|}[\f{x R}\nu]\right\}\
,\label{HnuR}\\ 
S_{out}[x, |n|,\nu]&=&-\f{\nu\left(1-\f 1{\nu^2}\right) p^{|n|}_{\mu_{x}}{}'
[\f 1\nu]I_{|n|\nu}[x] + x p^{|n|}_{\mu_{x}} [\f 1\nu]
I'_{|n|\nu}[x]}{\nu\left(1-\f 1{\nu^2}\right) p^{|n|}_{\mu_{x}}{}'[\f
1\nu]K_{|n|\nu}[x] + x p^{|n|}_{\mu_{x}} [\f 1\nu] K'_{|n|\nu}[x]}\ ,
\\  
S_{in}[x, |n|,\nu]&=&-\f{\nu\left(1-\f 1{\nu^2}\right) 
q^{|n|}_{\mu_{x}}{}'[\f 1\nu]K_{|n|\nu}[x]+ x q^{|n|}_{\mu_{x}}[\f 1\nu]
K'_{|n|\nu}[x]}{\nu\left(1-\f 1{\nu^2}\right)
p^{|n|}_{\mu_{x}}{}'[\f 1\nu]K_{|n|\nu}[x]+ x p^{|n|}_{\mu_{x}} [\f 1\nu]
K'_{|n|\nu}[x]}\ .
\en

In the last term in Eq.\Ref{HnuR} we used  
Eq.\Ref{Prop} and changed the argument of the Bessel functions from 
$xR$ to $xR/\nu$. By construction, the self-energy is a $C^1$-regular function 
at the surface of the string because Green's function is expressed in terms 
of the functions given in Eq.\Ref{HomSol}. The renormalization is done by 
subtracting the same function in the regions outside and inside the string. 
Therefore, the expression given previously which corresponds to the 
self-potential is a $C^1$-regular function at the string's surface mode by 
mode, which is more suitable for numerical simulations.  

First of all let us consider the self-energy for a particle situated outside
the string with $R\geq 1$. In the neighborhood of string's surface the
series converges very slowly. To simplify numerical calculations we 
represent it in the following form  
\bn
&&\f 1RL(\nu, R) = \f 2{\pi R}\int_0^\infty dx \left\{\nu
K_0[x]I_0[x]-K_0[\f x\nu]I_0[\f x\nu] + \nu
S_{out}[\f xR, 0,\nu]K^2_0[x]\right\}\label{LnuR}\\ 
&+&\f 4{\pi R}\sum_{n=1}^N n\int_0^\infty dx \left\{\nu
K_{n\nu}[nx]I_{n\nu}[nx]-K_{n}[\f{nx}\nu]I_{n}[\f{nx}\nu] + \nu
S_{out}[\f{nx}R, n,\nu]K^2_{n\nu}[nx]\right\}\nonumber \\
&+&\f 4{\pi R}\sum_{n=N+1}^{+\infty}n\int_0^\infty dx \left\{\nu
K_{n\nu}[nx]I_{n\nu}[nx]-K_{n}[\f{nx}\nu]I_{n}[\f{nx}\nu] + \nu
S_{out}[\f{nx}R, n,\nu]K^2_{n\nu}[nx]\right\}\nonumber\ .
\en
In the last term of the previous equation, let us use(for sufficiently 
great $N$) the uniform expansion for Bessel functions\cite{AbraSteg}
\bn
K_p[pz]&\approx& \sqrt{\f{\pi t}{2p}}e^{-p\eta}\sum_{k=0}^\infty (-p)^{-k}
u_k[t]\ ,\ I_p[pz]\approx\sqrt{\f{t}{2\pi p}}e^{p\eta}\sum_{k=0}^\infty p^{-k}
u_k[t]\ ,\\
K'_p[pz]&\approx& -\sqrt{\f{\pi}{2ptz^2}}e^{-p\eta}\sum_{k=0}^\infty (-p)^{-k}
\overline{u}_k[t]\ ,\ I'_p[pz]\approx\sqrt{\f{1}{2\pi
ptz^2}}e^{p\eta}\sum_{k=0}^\infty p^{-k}\overline{u}_k[t]\ ,\nonumber 
\en 
where 
\bn
t[z]&=&\f 1{\sqrt{1+z^2}}\ ,\ \eta[z] = \sqrt{1+z^2}+\ln\f z{1+\sqrt{1+z^2}}\
,\nonumber\\ 
u_k[t]&=&\f 12 t^2(1-t^2)u'_{k-1}[t]+\f 18 \int_0^t (1-5t^2)u_{k-1}[t]dt \
,\ u_0[t]=1\\ 
\overline{u}_k[t]&=&u_k[t]+t(t^2-1)\left\{\f
12u_{k-1}[t]+tu'_{k-1}[t]\right\}\ ,\ \overline{u}_0[t]=1\ , \nonumber
\en
and the uniform expansion of Legendre's functions found in
Ref.\cite{BordKhus} which have the form below
\bn
p^n_\mu[z]&=&\f 1{n!}\left[\f{1+\gamma^2
v^2}{1+\gamma^2}\right]^{\f 14}e^{nS}\sum_{k=0}^\infty n^{-k}\Pi_k[v]\
,\nonumber \\
q^n_\mu[z]&=&\f{(n-1)!}2\left[\f{1+\gamma^2 
v^2}{1+\gamma^2}\right]^{\f 14}e^{-nS}\sum_{k=0}^\infty
(-n)^{-k}\Pi_k[v]\ ,\nonumber\\
\f 1n\f d{dz}p^n_\mu &=&-\f 1{n!}\left[\f{1+\gamma^2
v^2}{1+\gamma^2}\right]^{\f
34}\f{1+\gamma^2}{1-v^2}e^{nS}\sum_{k=0}^\infty n^{-k}\overline{\Pi}_k[v]\
,\\ 
\f 1n\f d{dz}q^n_\mu &=&\f{(n-1)!}2\left[\f{1+\gamma^2 
v^2}{1+\gamma^2}\right]^{\f 34}\f{1+\gamma^2}{1-v^2}e^{-nS}\sum_{k=0}^\infty
(-n)^{-k}\overline{\Pi}_k[v]\ ,\nonumber  
\en
where 
\bn
\mu &=&-\f 12 +\f 12\sqrt{1-\left(\f{2nx}{\tan\epsilon}\right)^2}\ ,\ v=\f
z{\sqrt{1+\gamma^2(1-z^2)}}\ ,\ \gamma=\f x{\tan\epsilon}\ ,\nonumber \\
S&=&\f 12\ln\left[\f{1-v}{1+v}\f{1}{1+\gamma^2}\right] - \gamma (\arctan
[\gamma v] - \arctan [\gamma])\ ,\nonumber \\
\Pi_{k+1}[v]&=&\f{1-v^2}{2}\f{1+\gamma^2 v^2}{1+\gamma^2}\Pi'_k[v]
\nonumber \\
&-&\f{\gamma^2}{8(1+\gamma^2)}\int_1^vdv'\left\{5v'^2+\f 1{\gamma^2} - 1 -
\f{1+\gamma^2}{\gamma^2(1+\gamma^2v'^2)}\right\}\Pi_k[v']\ ,\ \Pi_0[v]=1\
,\\ 
\overline{\Pi}_k[v]&=&\Pi_k[v]-\f{\gamma^2v(1-v^2)}{2(1+\gamma^2)}\Pi_{k-1}[v]
-\f{(1-v^2)(1+\gamma^2v^2)}{1+\gamma^2}\Pi'_{k-1}[v]\ ,\
\overline{\Pi}_0[v]=1\ .\nonumber 
\en 
Then, taking into account these previous formulas we have the following 
expression for the last term in Eq.\Ref{LnuR}

\bn
&&\f 4{\pi R}\sum_{n=N+1}^{+\infty}n\int_0^\infty dx \left\{\nu
K_{n\nu}[nx]I_{n\nu}[nx]-K_{n}[\f{nx}\nu]I_{n}[\f{nx}\nu] + \nu
S_{out}[\f{nx}R, n,\nu]K^2_{n\nu}[nx]\right\}\nonumber \\
&&\approx\f{\nu^2-1}{4\pi\nu R}\sum_{n=N+1}^\infty\f 1{n^2} \int_0^\infty dz
\left\{-z^2t[z]^5(1-5t[z]^2)+\f{z^2}{R^2}t[z]t[\f zR]^4e^{-2n\nu
(\eta[z]-\eta[\f zR])}\right\}\ ,
\en
which may be expressed in terms of the function 
\be
\Phi(z,s,v)=\sum_{n=0}^\infty (v+z)^{-s}z^n\ . 
\ee
Using integral representation for this function\cite{BateErde} we obtain,
finally, the following expression for this term in Eq.\Ref{LnuR}
\bn
&&\f 4{\pi R}\sum_{n=N+1}^{+\infty}n\int_0^\infty dx \left\{\nu
K_{n\nu}[nx]I_{n\nu}[nx]-K_{n}[\f{nx}\nu]I_{n}[\f{nx}\nu] + \nu
S_{out}[\f{nx}R, n,\nu]K^2_{n\nu}[nx]\right\}\nonumber \\
&&\approx\f{\nu^2-1}{4\pi\nu R}\left\{\f 13\zeta_{H}[2,N+1] + \f 1{R^2}
\int_0^\infty dz z^2t[z]t[\f zR]^4\int_0^\infty dy 
y\f{e^{-N[y+2\nu(\eta[z]-\eta[\f zR])]}}{e^{y+2\nu(\eta[z]-\eta[\f
zR])}-1}\right\}\ ,\label{l}
\en
which is suitable for numerical calculations. 

We use the same approach for function $H(\nu,R)$ and obtain the 
following result for the series corresponding to function $H(\nu,R)$
\bn
&&\f{4}{\pi}\sum_{n=N+1}^{+\infty}n\int_0^\infty dx \left\{
p^{n}_{\mu_{x}}[x_{R}] \left(q^{n}_{\mu_{x}} [x_{R}]+S_{in}[x, n,\nu
]p^{n}_{\mu_{x}} [x_{R}]\right) - K_{n}[\f{x R}\nu] I_{n}[\f{x
R}\nu]\right\}\nonumber \\
&&\approx \f{\nu^2-1}{4\pi\nu}\left\{R\zeta_{H}[2,N+1]-
\int_0^\infty dz z^2t[zR]t[z]^4\int_0^\infty dy 
y\f{e^{-N[y+2(\tilde{\eta}[z,1]-\tilde{\eta}[z,R])]}}{e^{y+2(\tilde{\eta}[z,1]
- \tilde{\eta}[z,R])}-1}\right\}\ , \label{h}
\en
where  
\bn
\tilde{\eta}[z,R]&=&\ln\f{zR}{\sqrt{1+z^2R^2}+\sqrt{1-R^2\sin^2\epsilon}}
\nonumber \\
&-& \f z{\sin\epsilon}\left[\arctan\f{z
\sqrt{1-R^2\sin^2\epsilon}}{\sin\epsilon 
\sqrt{1+z^2R^2}}-\arctan\f z{\sin\epsilon}\right]\ , \\
\eta[z]&=&\sqrt{1+z^2}+\ln\f z{1+\sqrt{1+z^2}}\ ,
\en
and $\zeta_H(s,x)$ is the Hurwitz zeta function. At the surface of the
string $R=1$, and expressions given by Eqs.\Ref{l} and \Ref{h} coincide 
and are equal to 
\be
\f{\nu^2-1}\nu\f{\zeta_{H}[2,N+1]}{6\pi}\ .  
\ee
Therefore the self-energy is continuous at string's surface. For
numerical simulations we used previous formulas for $N=0$.  

In order to analyze the self-force near string's surface let us
consider more carefully the self-energy for a particle outside the
string. For $N=0$ we have 
\bn
\f 1RL(\nu, R) &=& \f 2{\pi R}\int_0^\infty dx \left\{\nu
K_0[x]I_0[x]-K_0[\f x\nu]I_0[\f x\nu] + \nu S_{out}[\f xR,
0,\nu]K^2_0[x]\right\}\\  
&+&\f\pi{72}\f{\nu^2-1}{\nu R} +  \f{\nu^2-1}{4\pi\nu}
\int_0^\infty dz z^2t[zR]t[z]^4\int_0^\infty dy 
\f y{e^{y+2\nu(\eta[zR]-\eta[z])}-1}\ .\nonumber
\en 
The last term will give a logarithmic divergence at string's
surface. In order to see this let us represent it in the form 
\be
\int_0^\infty \f{ydy}{e^{y+2\nu p}-1}=\int_{2\nu p}^\infty \f{ydy}{e^y -
1} + 2\nu p\ln (1-e^{-2\nu p})\ ,  
\ee 
where 
\be
p = \eta[zR]-\eta[z] =\ln R + \ln\f{\sqrt{1+z^2}+1}{\sqrt{1+z^2R^2}+1}
+\sqrt{1+z^2R^2} - \sqrt{1+z^2}\ . 
\ee
The derivative of the integral in rhs with respect to $R$ is finite at the
point $R=1$, but the derivative of the second term in rhs gives
a logarithmic divergence at $R=1$ because 
\be
\f{dp}{dR} = \f{\sqrt{1+z^2R^2}}R 
\ee
is finite at string's surface.  

Taking into account previous formulas, we obtain the following result which
are divergent at $R=1$
\bn
\f 1RL(\nu, R) &=& \f{\nu^2-1}8(R-1)\ln(R-1)+l(\nu, R)\ ,\ R>1\\
H(\nu, R)      &=& -\f{\nu^2-1}8(1-R)\ln(1-R)+h(\nu, R)\ ,\ R<1\ .
\en 
The functions $l(\nu, R)$ and $h(\nu, R)$ and their first derivatives with
respect to $R$ are finite at $R=1$.   

Taking the derivative of above expressions inside and outside the string
we find that the divergence at string's surface is
given by Eq.\Ref{Fdiv}. This expression does not depend on the number $N$
because this divergence appears associated with small $y$, but for small 
$y$ the integrand does not depend on $N$. Therefore each term in the 
expression for the self-force is finite and continuous at the string surface 
but the sum of the series is logarithmically divergent. 

\end{document}